\documentstyle[11pt]{article}

\begin{document}
\centerline{\large \bf Random Quantum Billiards}
\vskip 0.3cm
\centerline{Henrik J. Pedersen and A. D. Jackson}
\vskip 0.2cm
\centerline{The Niels Bohr Institute}
\centerline{Blegdamsvej 17, DK 2100 Copenhagen {\O}}
\centerline{Denmark}

\vskip 3.0cm

\centerline{\bf Abstract}

\vskip 0.4cm

We present a random matrix model suitable for the quantum mechanical 
description of a particle confined to move inside a two-dimensional 
domain.  Here, the ensemble average corresponds to an average over 
domain shapes.  Although this approach is formally distinct from that 
of the familiar Gaussian orthogonal ensemble, there is remarkable 
agreement in the resulting distributions of level spacings and 
in the spectral rigidity.  We discuss possible applications of this 
ensemble to the analysis of real data sets.

\newpage

\centerline{\bf I. Introduction}
\vskip 0.3cm
Received wisdom suggests that universal features of the spectrum of 
completely chaotic quantum billiards, such as the distribution of 
level spacings and the spectral rigidity, are identical to those of 
the Gaussian orthogonal ensemble (GOE).  Both experimental data [1,2] 
and the results of numerical studies [3,4,5] provide considerable support for 
this belief.  With a proper choice of the zero and the unit of the 
energy scale, the GOE has become the common standard of comparison.  
However, in spite of considerable analytical efforts [6,7,8,9], 
there is still no proof of the Bohigas conjecture [5] that the spectral 
correlations of quantum systems with chaotic classical analogues are 
given by random matrix theory.  Indeed, some reflection on the physical 
nature of the two problems suggests that this agreement is surprising.  
The GOE emerges from the ensemble of real symmetric random matrices in the 
limit of infinite dimension, $M$.  Each matrix, $H$, is drawn from a 
Gaussian distribution $\exp{[-{\rm Tr}(H^T H)/2]}$.  The average value 
of each matrix element is zero.  The quantum billiard problem requires the 
solution of the Schr{\"o}dinger equation for a particle which moves 
freely within a bounded domain.  The eigenvalues are typically obtained 
by expanding the problem in some suitable basis and diagonalizing a 
matrix of finite dimension.  Obviously, all eigenvalues of quantum 
billiards are positive.  This stands in contrast to the fact that 
the eigenvalues of the GOE are distributed symmetrically about $0$ 
according to the familiar semicircular law, $\rho(x) \sim \sqrt{2M 
- x^2}$ with support from $-\sqrt{2M}$ to $+\sqrt{2M}$.  The 
eigenvalues are not bounded from below in the thermodynamic limit.  
Since any given eigenvalue diverges as the dimension of the matrix is 
increased, questions associated with the convergence of eigenvalues and 
eigenvectors are largely without meaning.  This behaviour is again 
qualitatively different from that of real quantum billiard problems 
for which asymptotic power law convergence is expected for both 
eigenvalues and eigenfunctions.  Further, the GOE is invariant under 
orthogonal transformations, and there is thus no preferred basis.  
While this represents a considerable technical simplification,
it also indicates another difference from quantum billiards.  There, we have 
the expectation of increasing nodal complexity with increasing energy and, 
thus, a qualitative basis preference.   

In spite of these remarks, the intention of this paper is not to advocate 
rejection of the GOE for the analysis of quantum billiard problems.  Rather, 
we believe it to be desirable to consider the properties of a new 
random matrix ensemble which more faithfully imitates the properties 
of quantum billiard problems by respecting the positivity of the 
spectrum and maintaining the qualitative basis preference just mentioned.  
(We note that the properties of finite ensembles of real ``random'' 
billiards have been studied previously.  See, e.g., refs.\,[10] and [11].)
We will show numerically that the spectral correlations of this 
model are identical to those of the GOE.

The model and its physical motivation will be introduced in section II.  
In section III we will consider the accuracy of eigenvalues calculated with 
this model and demonstrate power law convergence in eigenvalues and 
eigenfunctions with increasing matrix dimension.  In section IV we  
present the level spacing distribution and spectral rigidity for our 
random billiard ensemble (RBE).  These results will be found to be 
statistically indistinguishable from the GOE.  Although we will concentrate 
on numerical investigation of the RBE, it has been constructed with 
sufficient simplicity that its analytic investigation should also be 
tractable.  In section V we will compare the wave functions obtained from 
this model with those of the GOE and of an ensemble of real quantum 
billiards.  Discussion and conclusions will be relegated to section VI. 
     
\vskip 0.5cm
\centerline{\bf II. The Construction of the Model}
\vskip 0.3cm     
Our random matrix model of quantum billiards will be patterned on 
the standard conformal mapping method for the solution of the 
Schr{\"o}dinger equation 
for a point particle in confined to the interior of a simply-connected 
domain in the two-dimensional $uv$-plane [12,13].  It is always possible to 
find a new pair of variables, $x$ and $y$, such that the system is mapped 
onto some other convenient simply-connected domain, e.g., the unit 
disk, for which the Laplacian is separable and for which its 
eigenfunctions, $\phi_i (x,y)$, and eigenvalues, $\epsilon_i$, are known.  
(We order the states so that $\epsilon_{i+1} \ge \epsilon_i$.  Evidently, 
$\epsilon_i > 0$ for all $i$.  The $\phi_i$ are orthogonal and normalized 
in the $xy$-plane on the weight $1$.)  The condition of simple-connectedness 
is sufficient to ensure that this mapping will be 
conformal with 
\begin{equation}
u + i v = w (z) \ ,
\label{e1}
\end{equation}
where $z = x + i y$.  We now expand the desired eigenfunctions, 
$\Phi (u,v)$, in the complete basis $\phi_i$ with coefficients $\Phi_i$.  
Since the transformation is conformal, the matrix elements of the 
Hamiltonian are particularly simple:
\begin{eqnarray}
H_{ij} & = & \int \, du \, dv \, \phi_i (x,y) \left( - \nabla^2_w \right) 
\phi_j (x,y) \nonumber \\
    {} & = & \int \, dx \, dy \, \phi_i (x,y) \left( 
- \nabla^2_z \right) \phi_j (x,y) = \epsilon_i \delta{ij} \ \ .
\label{e2}
\end{eqnarray}
The Hamiltonian matrix is thus diagonal with elements $\epsilon_i$ which 
are {\em independent\/} of the shape of the initial domain.  As a 
consequence of the separability of the Laplacian in $z$, the spacings 
between adjacent levels, $\epsilon_{i+1} - \epsilon_i$, will have a Poisson 
distribution.  The average level spacing will be independent of energy for a 
two-dimensional quantum billiard.  Since the transformation 
of variables destroys the orthogonality of the $\phi_i$, it is necessary 
to consider the overlap integrals
\begin{equation}
N_{ij} = \int \, du \, dv \, \phi_i (x,y) \phi_j (x,y) = 
         \int \, dx \, dy \, | dw /dz |^2 \, 
                             \phi_i (x,y) \phi_j (x,y) \ \ .
\label{e3}
\end{equation}     
The Schr{\"o}dinger equation now assumes the form 
\begin{equation}
\epsilon_i \Phi_i = \lambda N_{ij} \Phi_j \ \ .
\label{e4}
\end{equation}
All information regarding the shape of the initial domain is carried 
by the matrix $N$.  One physical property of $N$ is worth noting.  
Assume that the mapping is in some sense smooth. The functions $\phi_i$ will oscillate rapidly on a scale set by 
the mapping for sufficiently large $i$.  Given the normalization of 
the $\phi_i$, the diagonal 
elements, $N_{ii}$, will then approach the area of the original domain, 
which can be set equal to $1$. 

We can express the elements in $N$ in a more suggestive fashion.  Since 
the $\phi_i$ are complete, we can use the Gram-Schmidt procedure to 
construct a basis of states, $\psi_i$, which are orthonormal on the 
weight $|dw/dz|^2$.  If we write
\begin{equation}
\phi_i = \sum_{\ell=1}^{i} \, r_{i \ell} \psi_{\ell} \ \ ,
\label{e5}
\end{equation}
the elements of $N$ are simply 
\begin{equation}
N_{ij} = {\vec r}_i \cdot {\vec r}_j \ \ .
\label{e6}
\end{equation}  
This structure of $N$ and the positivity of the $\epsilon_i$ are 
sufficient to ensure the positivity of the eigenvalues for every 
domain.  The physical constraint on the diagonal elements of $N$ assumes 
the form ${\vec r}_i \cdot {\vec r}_i \rightarrow 1$ as $i \rightarrow 
\infty$ for a billiard of unit area. 

Everything said so far has been exact.  We now use the form of 
this calculation to design a random matrix model of quantum billiards:
\begin{quotation}
Draw $\epsilon_1$ and all subsequent level spacings, 
$\epsilon_{i+1} - \epsilon_i$, at random on a Poisson distribution (with 
variance $1$).  Draw the elements $r_{i \ell}$ for $\ell \le i$ at random 
on, e.g., a Gaussian distribution with mean value $0$ and variance 
$1/i$ so that ${\vec r}_i \cdot {\vec r}_i = 1$ on average.  Use 
these quantities to construct the Schr{\"o}dinger equation, eqn.(4).  
\end{quotation}
This defines our random billiard ensemble (RBE).  The ensemble average 
over the ${\vec r}_i$ corresponds to an average over billiard shapes.  
A further average over the $\epsilon_i$ corresponds to averaging over 
the shape (and basis) chosen in the $z$ plane.  Thus, this average is 
not necessary but can be convenient.

Several comments are in order.  First, every matrix selected according 
to this algorithm will have a positive spectrum.  Second, it is simple 
and meaningful to increase the dimension of the matrix in order to study 
the convergence properties of individual eigenvalues.  (Draw a single new 
${\vec r}_{M+1}$ and a new $\epsilon_{M+1}$ according to the above rules 
leaving all preceding ${\vec r}_i$ and $\epsilon_i$ unaltered.)  
As we shall see below, eigenvalues (and eigenfunctions) will show 
power law convergence for large $M$.  The usual numerical (and experimental) 
approach to billiard problems is to consider many levels for a given 
shape.  This can lead to the diagonalization of inconveniently large 
matrices.  Here, greater economy is possible.  It is possible to obtain 
impressive statistical accuracy for the usual measures of this ensemble 
(e.g., the level spacing distribution or the spectral rigidity) by considering 
a modest number of eigenvalues for each of a large ensemble of ``shapes''.  
As we shall see in section IV, the level spacing distribution, 
$e_{k+1} - e_k$, converges rapidly (with $k$) to its asymptotic form.  

Finally, it is useful to perform a rough counting of the number of free 
parameters in order to get a sense of those features of real billiards 
which have been lost in the RBE.  For a matrix of dimension $M$, 
the matrix $N$ contains ${\cal O}( M^2 )$ free parameters.  Since the 
Laplacian is assumed to be separable in the coordinates chosen in $z$, 
the $M$ basis states $\phi_i$ are the various products of two of the 
$\sqrt{M}$ independent functions of one variable each.  The overlap matrix 
resulting from the weight function of a conformal map, $|dw/dz|^2$, should 
thus contain roughly ${\cal O}(\sqrt{M})$ free parameters.  A local (but 
not conformal) map would lead to ${\cal O}(M)$ free parameters.  By 
extension, our random matrix model corresponds roughly to 
choosing a positive but non-local weight function.  This is, of course, 
not consistent with the 
assumption of the diagonality of $H$, which is correct only when the 
mapping is conformal.  Such inconsistency is a hallmark of random matrix 
models and is not necessarily cause for alarm.

\vskip 0.7cm
\centerline{\bf III. Convergence of the Calculations}
\vskip 0.5cm

Before reporting the results of this model, we wish to discuss the 
convergence of calculated eigenvalues and eigenfunctions as a function 
of the dimension of the matrix, $M$.  Given the model of section II, we 
expect to find (fully converged) energies of $e_k \approx k$ and   
an average level spacing of $1$ (at all energies).   Only fully converged 
eigenvalues are of physical interest, and these will represent a small 
fraction of the eigenvalues of any given matrix.  Nevertheless, useful 
information about convergence is provided by looking at the change in 
$e_k$ as $M \rightarrow M+1$ for a broad range of $k$.  The logarithm of 
this change (averaged over ensembles of matrices ) as a function of $k/M$ 
for $M = 200$, $500$, $700$, and $1000$ is shown as the heavy band in 
fig.\,1.

\begin{figure}[here]
\input epsf
\epsfxsize=10cm
\epsfysize=10cm
\centerline{
\epsfbox{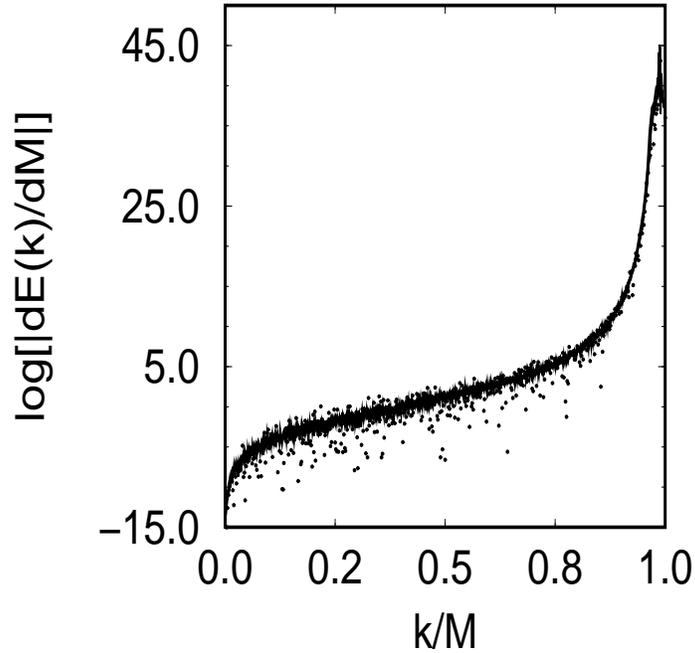} }
\caption{ Plot of $de_k /dM$ as a function of $k/M$.  The 
heavy band represents results for $M = 200$, $500$, $700$, and $1000$.  
The points are the results for a single matrix with $M=500$. }
\end{figure}

The surprise here is that $de_k^{(M)}/dM$ is a function 
only of the ratio $k/M$.  While there are considerable fluctuations about 
this curve for the eigenvalues of single matrices, these are not so large as 
to invalidate this average curve.  This is indicated by the points in 
fig.\,1, which represent the results for a single matrix of dimension 
$M=500$.  The data reveal a region of small $k/M$ (in which the energies 
are accurate) where $de_k^{(M)}/dM$ is proportional to $(k/M)^2$.  
This asymptotic behaviour 
is followed by a region in which $de_k^{(M)}/dM$ increases roughly 
expontentially (and in which the energies are not of useful accuracy).  
The results of this figure can be integrated to construct the average 
error in $e_k$ due to finite matrix size.  Clearly, the form of the 
energy will be given as 
\begin{equation}
e_k^{(M)} = e_k + M f(k/M).
\label{e7}
\end{equation} 
The data show that $f(k/M) \sim -(k/M)^{2}$ in the useful limit of small 
$k/M$.  This means that we can obtain $k$ levels which meet any 
fixed accuracy criterion by allowing $M$ to grow like $k^2$.  

It is possible to understand this behaviour by re-expressing eqns.(4) 
and (6) in the orthonormal basis of eqn.(5).
\begin{equation}
h \Psi = R^{T} \frac{1}{H} R \Psi = \frac{1}{\lambda} \Psi \ \ , 
\label{e8}
\end{equation} 
where $R$ is the lower triangular matrix whose elements are the $r_{i \ell}$.
The eigenvalues and eigenfunctions of eqn.(8) are identical to those of 
eqns.(4) and (6).  Given the structure of $R$ and the diagonality of $H$, 
we can write $h$ (in the Gram-Schmidt basis) as  
\begin{equation}
h_{ij}^{(M)} = \sum_{k={\rm max}\{i,j\}}^M \ 
\frac{r_{ki} r_{kj}}{\epsilon_k} \ \ .
\label{e9}
\end{equation}
Imagine that we have determined the eigenvalues and eigenfunctions of 
$h^{(M-1)}$.  Working in this basis, we consider the change in 
$\delta h = h^{(M)} - h^{(M-1)}$, which results from increasing the 
dimension by $1$.  Every element of $\delta h$ will be of  ${\cal O}(M^{-2})$.  Since the 
difference between $1/e_{k}^{(M-1)}$ and $1/e_{M}^{(M-1)}$ (which equals 
$0$) will be ${\cal O}(1/k)$ for $M >> k$, it is legitimate to estimate 
the change in eigenvalues using perturbation theory.  Making use of the 
fact that $\epsilon_k \approx k$ on average, zeroth-order perturbation theory yields
\begin{equation}
\left\langle \frac{1}{e_k^{(M)}} \right\rangle  - 
\left\langle \frac{1}{e_k^{(M-1)}} \right\rangle = \frac{1}{M^2} 
\label{e10}
\end{equation}     
when an ensemble average is performed over the elements of $\delta h$.  
Summing these changes over $M$ yields the consistent result that 
$\langle 1/e_k \rangle \rightarrow 1/k$ as $M \rightarrow \infty$ 
and that $\langle e_k \rangle$ will be approximately $k$ as expected.  
Making the tentative replacement of $\langle 1/e_k \rangle$ by 
$1/\langle e_k \rangle$, this suggests that 
\begin{equation}
\langle de_k / dM \rangle \approx -  \frac{k^2}{M^2} \ \ ,
\label{e11}
\end{equation}
which is the asymptotic result of fig.\,1. We note that it is possible to
provide a rigorous proof, valid for all $(k/M)$, of the scaling behaviour
indicated in fig.1 and eqn.(7). This will be described elsewhere.   

The ensemble average of every off-diagonal matrix element of $\delta h$ 
is $0$, and the variance is $1/M^4$.  Thus, first-order perturbation 
theory then tells us that the probability of finding the state $\psi_M$ 
in the $k$-th eigenfunction should be proportional to $k^2/M^4$ for $M >> k$. 
(Given the symmetry of first-order perturbation theory, this is also the 
probability of finding the state $\psi_k$ in the $M$-th eigenfunction for 
$k << M$.)  We shall see in the next section that this behaviour of 
eigenfunctions is correct.  

Since the dominant effect of increasing the size of the basis can 
be estimated from zeroth-order perturbation theory, it is easy to 
accelerate the convergence of these calculations significantly.  
Specifically, the upper limit of the sum in eqn.(9) can be extended 
to some number much larger than the size of the largest matrix 
to be considered.  This matrix can be truncated to dimension $M-1$ as 
before.  Now, the effect of increasing the dimension to $M$ now is 
the addition of a single row and column.  The original matrix of 
dimension $M-1$ is unaltered, and zeroth-order perturbation theory 
will not change the energy of state $k$.  To the extent that this 
eigenfunction is dominated by a single state in the Gram-Schmidt basis, 
first-order perturbation theory gives a correction of order $(k/M^{3})$ to 
$1/\langle e_k\rangle $.  This is to be compared with the result of eqn.(10) above.  
Except where otherwise noted, we have use this method to accelerate 
the convergence of the numerical calculations reported below.)  

The above estimates indicate slow but adequate (power law) convergence 
of eigenvalues as a function of $M$ and completely acceptable convergence 
of eigenfunctions.

We note that the combination of power law and exponential behaviour 
indicated in fig.\,1 makes it easy to draw misleading conclusions regarding 
the convergence of these calculations.  Specifically, study of the error 
for fixed $k$ and large $M$ would reveal only the explicit power-law 
dependence on $M$.  The $M$ dependence of the exponential is easily missed.  
However, study for fixed $M$ as a function of $k$ (relatively small) readily 
reveals an exponential divergence with $k$.  The naive combination of these 
results can lead to the erroneous conclusion that the size of the matrix 
must grow exponentially with $k$ in order to meet a fixed accuracy criterion. 
As already indicated, this is not correct.

\vskip 0.7cm
\centerline{\bf IV. Level Densities, the Level Spacing Distribution,}
\centerline{\bf and the Spectral Rigidity} 
\vskip 0.5cm

In this section, we wish to present a number of numerical results for 
the random billiard ensemble introduced in the section II.  These include 
the one-body level density, the level spacing distribution, and the 
spectral rigidity, $\Delta_3 (L)$.  Fig.\,2 shows the one-body level 
density obtained as an average over $300$ matrices of dimension $M = 300$.  
\vskip 2cm
\begin{figure}[here]
\input epsf
\epsfxsize=10cm
\epsfysize=10cm
\centerline{
\epsfbox{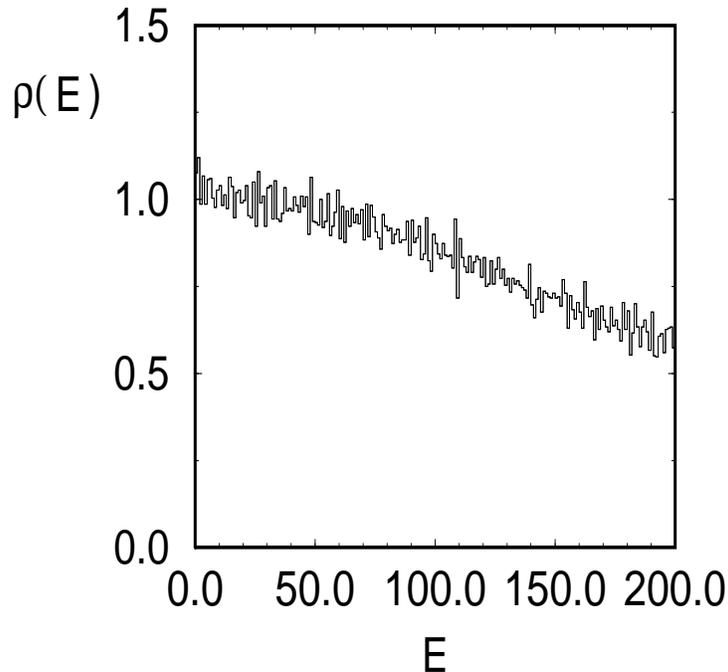} }
\caption{The level density for the RBE as obtained for $300$ 
matrices of dimension $M=300$. }
\end{figure}

Given the error estimates of the preceding section, we expect that only the 
first $50$ states (with energies from $0$ to roughly $50$) are 
quantitatively useful.   The resulting density of states is nearly constant 
over this limited energy range and equal to $1$, as expected for a family 
of two-dimensional billiards of unit area.  As indicated in section III, 
this range of constant level density grows as $M$ increases.  The one-body 
level density for larger energies has necessarily been constructed from 
eigenvalues which are not of quantitative accuracy.  As expected, the 
average spacing between these states is too large and the average level 
density is less $1$. 

In order to combine data for level spacing distributions from various parts 
of the spectrum, it is first necessary to scale the energies with the local 
average level spacing (i.e., to ``unfold'' the spectrum).  This is not 
essential here since we always work with converged eigenvalues and since 
the average level spacing is always very close to $1$ for the levels 
considered.  Nevertheless, we performed such a scaling by first binning 
pairs of levels according to their mean energy and then determining the 
average level spacing within each bin.  Since our model averages over 
billiard shapes, it is easy to consider the distribution of level spacings 
between states $k+1$ and $k$ for any fixed value of $k$.  Level spacing 
distributions obtained for small $k$ contain unwanted edge effects.  (For 
example, the distribution for $k=1$ shows a significantly enhanced 
probability for small level spacings.)  Fortunately, the level spacing 
distribution reaches a stable asymptotic form rapidly with increasing 
$k$.  Thus, fig.\,3 shows the RBE level spacing distribution of $1.5 \times 10^5$ 
levels with $19 \le k \le 30$ obtained from matrices with $M = 200$.  We 
also show the exact level spacing distribution for the GOE.  (The familiar 
Wigner surmise is not adequate for our purposes.)  Cursory inspection 
suggests a fully quantitative fit. 

\begin{figure}[here]
\input epsf
\epsfxsize=10cm
\epsfysize=10cm
\centerline{
\epsfbox{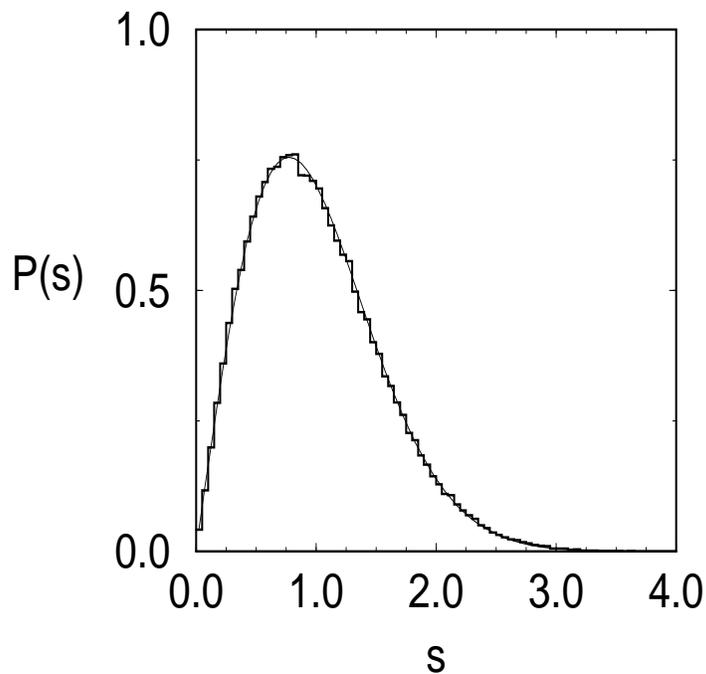} }
\caption{The histogram shows the RBE level spacing distribution 
of $1.5 \times 10^5$ level spacings using levels $k=19$ to $30$ obtained 
from the diagonalization of matrices of dimension $200$.  The solid line 
represents the exact level spacing distribution of the GOE. }
\end{figure}

This conclusion is supported by the 
average value of $\chi^2 = 1.28$ per datum over the range $0 \le s \le 4.0$. 
(This value of $\chi^2$ can be made closer to $1$ by diagonalizing larger 
matrices and using larger values of $k$.)  As a measure of the sensitivity 
of this comparison, we note that attempting to fit our data with the 
Wigner surmise leads to a $\chi^2$ of $4.6$ per datum.  Given that our data 
sample includes $80$ bins, this represents a remarkably strong preference 
for the exact GOE level spacing distribution.     

A somewhat longer-range characterization of the spectrum is provided 
by the spectral rigidity, $\Delta_3 (L)$, which is a measure of the 
fluctuations of a given sequence of unfolded levels about the best 
straight line fit [14].  Consider the level density for a given sequence of 
levels $x_i$ (with average spacings $1$) which lie in the range $-L/2 
\le x \le +L/2$:
\begin{equation}
\rho (x) = \sum \delta (x - x_i ) \ \ 
\end{equation}  
Using this density, define the ``staircase'' function
\begin{equation}
{\cal N}(x) = \int_{-L/2}^x \, dx' \, \rho (x') \ \ .
\end{equation}
Determine the mean square deviation relative to the best straight line fit 
for each such sequence and perform an ensemble average to obtain 
$\Delta_3 (L)$ as  
\begin{equation}
\Delta_3 (L) = \langle {\rm min} \, \frac{1}{L} \int_{-L/2}^{+L/2} \, 
dx \, ({\cal N}(x) - A - Bx)^2  \rangle \ \ ,
\end{equation}
where the minimization is with respect to the constants $A$ and $B$.
We have constructed $\Delta_3 (L)$ for our model of random quantum 
billiards over the range $0 \le L \le 30$.  The results shown in fig.\,4 
were obtained from an ensemble of $500$ matrices of dimension $300$ 
using levels $16$ to $60$.

\begin{figure}[here]
\input epsf
\epsfxsize=10cm
\epsfysize=10cm
\centerline{
\epsfbox{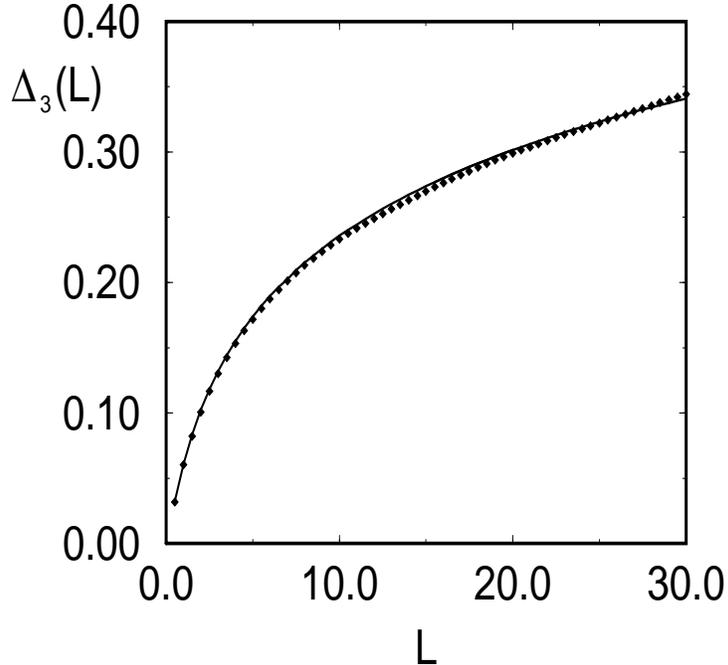} }
\caption{ The points show $\Delta_3 (L)$ for the RBE using 
levels $k=16$ to $60$ obtained from $500$ matrices of dimension $M=300$. }
\end{figure}

Again, we find quantitative agreement with 
the GOE; the small visible differences can be eliminated by working 
with higher levels (and larger matrices).  In both cases, $\Delta_3 (L)$ 
approaches ${\rm ln}{(2 \pi L)}/\pi^2$ 
in the limit of large $L$.  The correlations inherent in these spectra 
result in fluctuations which are much smaller than those of a random 
spectrum (with  a Poisson distribution of level spacings) for which 
$\Delta_3 (L) = L/15$.

\vskip 1.7cm
\centerline{\bf V. Eigenfunctions and Real Billiards}
\vskip 0.5cm

Having seen such striking agreement between RQB and GOE results, it is 
of interest to find some differences.  The most natural place to seek 
them is in the structure of eigenfunctions.  Thus, we consider the 
expansion coefficients of a given eigenfunction in the orthonormal basis 
of states defined in eqn.(5), 
\begin{equation}
\Psi_k = \sum \ {d_{\ell}}^{(k)} \, \psi_{\ell} \ ,
\label{e14}
\end{equation}
so that the $[{d_{\ell}}^{(k)}]^2$ can be regarded as probabilities.  As 
discussed in section III, we expect that the ensemble average of 
$[{d_{\ell}}^{(k)}]^2$ should grow like $\ell^2$ for $\ell << k$ and 
vanish like $1/ \ell^4$ for $\ell >> k$.  This expectation is confirmed 
by fig.\,5, which shows the $[{d_{\ell}}^{(k)}]^2$ for eigenvector $k=10$ 
averaged over $200$ matrices of dimension $400$.  (Here, we have not 
accelerated the convergence of the calculation.)  As expected, this 
result stands in sharp contrast 
to the eigenfunctions of the GOE.  Given the basis-independence of the GOE, 
the ensemble averaged probability should be precisely $1/N$ for each 
state.  One might expect to find some small variation if the diagonal 
elements of these matrices are first ordered in increasing value.  There 
is little sign of such dependence for states in the middle of the spectrum.  

Having found a difference between RQB and the GOE, it is appropriate to 
consider the behaviour of real billiards.  Any real billiard will have a 
weight function which will be smooth on some suitably small distance scale.  
Above, this led us to expect that the diagonal elements $N_{ii}$ will 
approach the area of the billiard in the limit of large $i$.  (This 
physically motivated constraint was built into our model so that the ensemble 
average of $r_{i \ell}^2$ is $1/i$ (for $\ell < i$).  Local smoothness of 
the weight has other consequences.  When the (two-dimensional) quantum numbers 
of states $\phi_i$ and $\phi_j$ are different, their product will 
oscillate over the domain of integration.  The magnitude of $N_{ij}$ 
should thus decrease with some power of the mismatch in the quantum numbers 
of the states.  Thus, we expect that the $[d_{{\ell}}^{(k)}]^2$ for real 
billiards should decrease when $k$ is very different from $\ell$ (and the 
two-dimensional quantum numbers are necessarily different).  The hull of 
maximum probabilities should show a power law decay away from the peak.  Given 
the fact that two states of approximately the same energy can have grossly 
different quantum numbers, the actual probabilities should show large 
local fluctuations.  

In order to demonstrate these effects, we have consider an ensemble of 
randomly selected real billiards.  Specifically, we have mapped a simply connected domain 
of unit area onto a square of length $1$ with one corner at the origin.  
This problem can be treated using the conformal mapping method of section I, 
and the basis functions are now simple sines.  All information regarding 
the original shape now resides in the analytic function $dw/dz$, which 
we have arbitrarily chosen to be a polynomial of order $3$ to $6$.  This 
polynomial is determined by its zeros which must lie outside the unit square.  
Subject to this constraint, we have drawn these zeros at random uniformly on a 
square of side $2.6$ centered at the origin and performed an ensemble average 
over $200$ real billiards.  (There is, on the one hand, no guarantee that 
the periodic orbits for these billiards are always of measure zero.  On 
the other, there is no assurance that the corresponding randomly selected 
mappings are univalent.  These matters should not be of importance for our 
limited considerations.)  The appropriate Gram-Schmidt basis was constructed 
for each shape drawn, and the $[{d_{\ell}}^{(k)}]^2$ were determined.  
The value of the $[{d_{\ell}}^{(k)}]^2$ for eigenvector $k=10$ is also 
shown in fig.\,5.  (The eigenvectors were obtained from the 
diagonalization of matrices of dimension $300$.)  
The hull of the maximum probabilities (as well as the behaviour of the 
local geometric mean probability) is similar to our RBE results.  
We note, in particular, the power-law rise to the peak and a decay which 
is somewhat faster than the $1/k^4$ behaviour of fig.\,5.  (If required, 
this decay could be simulated more accurately in the RBE by drawing the 
elements $r_{ki}$ with a variance which decreases smoothly as $i$ ranges 
from $k$ to $1$.)  

\begin{figure}[here]
\input epsf
\epsfxsize=10cm
\epsfysize=10cm
\centerline{
\epsfbox{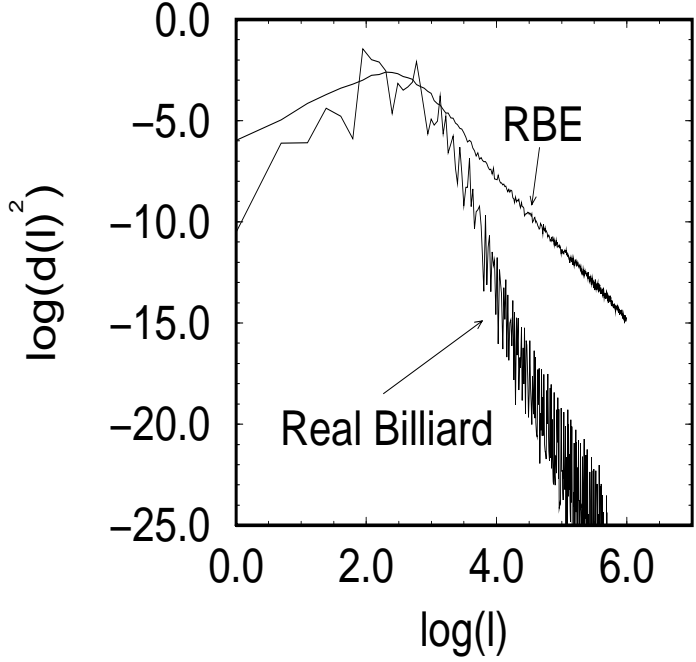} }
\caption{ Average probabilities, $[{d_{\ell}}^{(k)}]^2$, for the 
state $k=10$.  The upper curve was obtained from the RBE using $200$ matrices 
of dimension $M=400$.  The lower curve was obtained for the ensemble of 
$200$ real billiards as described in the text.
 }
\end{figure}

This figure also reveals the expected large local fluctuations which persist 
in spite of the ensemble average.  (Basis states with a significant mismatch 
in quantum numbers will tend to have small probabilities independent of 
the details of the mapping.)  It would, of course, be possible to construct 
a random matrix model which mimics these violent local fluctuations.  However, 
the underlying physical origin of these fluctuations (i.e., similarities 
or gross differences in the nodal structure of the basis functions) implies 
the existence of global correlations between matrix elements.  These are 
beyond the scope of simple random matrix models.  We believe that the 
present model is as successful in reproducing the average behaviour of 
real billiard wave functions as can reasonably be expected.

It is possible to study these differences in wave functions empirically 
by considering the parametric motion of energy levels.  Experimentally, 
one studies eigenvalues as some external variable (e.g., the temperature 
or the shape of the billiard) is varied continuously [15].  Measurements of 
the velocity autocorrelation function have already been reported in the 
literature [16], and curvature distributions are anticipated shortly [17].  
In the 
GOE, for example, one considers the eigenvalues of matrices $\cos{\lambda} H_1 
+ \sin{\lambda} H_2$ with $H_1$ and $H_2$ elements of the GOE.  Thus, it 
is meaningful to consider $d e_i / d\lambda$.  An ensemble average over 
$H_2$ (for any choice of $H_1$ and $\lambda$) leads to a Gaussian distribution 
of slopes with a variance which is independent of $H_1$ and $\lambda$.  
A similar calculation in the RBE would involve the replacement of the matrix 
$R$ of eqn.(8) by $\cos{\lambda} R_1 + \sin{\lambda} R_2$.  The ensemble 
average over $R_2$ also gives a Gaussian distribution of slopes.  In this 
case, however, the variance depends on $R_1$ and $\lambda$.  (This is 
also true for an ensemble of real billiards.)  This represents a 
clear distinction between RBE (and real billiards) and the GOE.  It 
can be studied with existing data.   What remains to be seen is whether 
the distribution of these variances is sufficiently general to be of 
interest. 

\newpage
\vskip 0.7cm
\centerline{\bf VI. Discussion and Conclusions}
\vskip 0.5cm

Despite considerable experimental and numerical success, there are no 
compelling arguments establishing a relationship between the properties 
of real quantum billiard systems and random matrix ensembles.  This led 
Ott to observe [18] that ``[w]hile there are some suggestive theoretical 
results supporting the random matrix conjecture for quantum chaos, \ldots the 
validity of this conjecture and its range of applicability, if valid, 
remain unsettled.''  In our view, the most disturbing aspects of the 
Gaussian orthogonal ensemble are (i) that its spectrum is unbounded 
from below, (ii) that, as a consequence, there is little meaning in 
considering the convergence of individual eigenvalues or eigenfunctions, 
and (iii) that the technical simplification of basis independence is 
physically unreasonable.  We have presented a random matrix model of 
quantum billiards with a much clearer connection to genuine billiard 
problems which meets each of these concerns.  The spectrum is 
necessarily positive.  Eigenvalues and eigenfunctions show power-law 
convergence as the dimension of the matrix is increased.  None the 
less, numerical investigation of the usual spectral measures (i.e., the 
level spacing distribution and the spectral rigidity) are in fully 
quantitative agreement with GOE results.  In this sense, the RBE can 
be thought of as an ``interpolating'' ensemble which serves to make 
the connection between real billiards and the GOE more plausible.  We 
believe that it represents significant new evidence in support of the 
random matrix conjecture for quantum chaos.

The most striking differences between the present results and those 
of the GOE are in the behaviour of eigenfunctions.  Here, our model 
is in good agreement with results obtained for an ensemble average 
over real quantum billiards.  We have noted that existing experimental 
data (e.g., the parametric motion of individual levels) can reveal 
shortcomings of the GOE.  It remains to be seen whether ``more realistic'' 
random matrix models such as the one considered here can extend the 
range of applicability of random matrix techniques in quantum chaos.

\vskip 0.4cm
\centerline{\bf Acknowledgements}
\vskip 0.3cm

We would like to thank Henrik Bruus, Thomas Guhr, Ronnie Mainieri, and 
Andreas Wirzba for countless discussions, helpful advice, and 
encouragement.

\vskip 0.7cm
\newpage
\centerline{\bf References}
\vskip 0.5cm

\noindent [1] C. Ellegaard, T. Guhr, K. Lindemann, H.Q. Lorensen, 
J. Nyg{\aa}rd, and M. Oxborrow, Phys.Rev.Lett.\ {\bf 75} (1995) 1546.

\vskip 0.2cm

\noindent [2] H.-D. Gr{\"a}f, H.L. Harney, H. Lengeler, C.H. Lewenkopf, 
C. Rangacharyulu, A. Richter, P. Schardt, and H.A. Weidenm{\"u}ller, 
Phys.Rev.Lett.\ {\bf 69} (1992) 1296.

\vskip 0.2cm

\noindent [3] M.V. Berry and M. Tabor, Proc.Roy.Soc.\ London 
{\bf A356} (1977) 375.

\vskip 0.2cm

\noindent [4] S.W. McDonald and A.N. Kaufman, Phys.Rev.Lett.\ {\bf 42} 
(1979) 1189.

\vskip 0.2cm

\noindent [5] O. Bohigas, M. Giannoni, and C. Schmidt, Phys.Rev.Lett.\ 
{\bf 52} (1984) 1; O. Bohigas and M. Giannoni, Lecture Notes in 
Physics {\bf 209} (1984) 1.

\vskip 0.2cm

\noindent [6] M.V. Berry, Proc.Roy.Soc.\ {\bf A400} (1985) 229.

\vskip 0.2cm

\noindent [7] E.B. Bogomolny and J.P. Keating, Phys.Rev.Lett.\ 
{\bf 77} (1996) 1472.
 
\vskip 0.2cm

\noindent [8] A.V. Andreev, O.Agam, B.D. Simons, and B.L. Altshuler, 
Nucl.Phys.\ {\bf B482} (1996) 536.

\vskip 0.2cm

\noindent [9] A. Altland and M. Zirnbauer, Phys.Rev.Lett.\ {\bf 77} 
(1996) 536.
 
\vskip 0.2cm 

\noindent [10] H. Bruus and A.D. Stone, Phys.Rev.\ {\bf B50} (1994) 
18275.

\vskip 0.2cm

\noindent [11] H. Bruus, C.H. Lewenkopf, E.R. Mucciolo, 
Phys.Rev.\ {\bf B53} (1996) 9968.

\vskip 0.2cm

\noindent [12] M. Robnik, J.Phys.A.\ {\bf 17} (1984) 1049.

\vskip 0.2cm

\noindent [13] M.V. Berry and M. Robnik, J.Phys.A.\ {\bf 19} (1986) 
649.

\vskip 0.2cm

\noindent [14] M.L. Mehta, ``Random Matrices'', 2nd.\ ed., Academic 
Press (Boston) 1991.

\vskip 0.2cm

\noindent [15] B.D. Simons and B.L. Altshuler, Phys.Rev.\ {\bf B48} 
(1993) 5422.

\vskip 0.2cm

\noindent [16] J.A. Folk, S.R. Patel, S.F. Godijn, A.G. Huibers, 
S.M. Cronenwett, C.M. Marcus, K. Campman, and A.S. Gossard, 
Phys.Rev.Lett.\ {\bf 76} (1996) 1699.

\vskip 0.2cm

\noindent [17] C. Ellegaard, private communication.

\vskip 0.2cm

\noindent [18] E. Ott, ``Chaos in Dynamical Systems'', Cambridge University 
Press (Cambridge) 1993, p.\,342.

\vskip 0.7cm
%
%
%
%
%
%
%
%
%
%
%
%
\end{document}